\documentclass[preprint,review,12pt]{elsarticle}

\usepackage{amssymb}
\usepackage{array}  
\usepackage{amsmath}
\usepackage{graphicx,color}

\usepackage{tikz}
\usepackage{natbib}
\journal{Mechanical Systems and Signal Processing}

\journal{Mechanical Systems and Signal Processing}

\begin{document}

\begin{frontmatter}
\title{A design methodology for nonlinear oscillator chains enabling energy localization tuning and soliton stability enhancement with optimal damping}

\author[label1]{A. Barbosa\corref{cor1}}
\ead{arthur.barbosa@femto-st.fr}
\cortext[cor1]{Corresponding author}
\author[label1]{N. Kacem}
\ead{najib.kacem@femto-st.fr}
\author[label1]{N. Bouhaddi}
\ead{noureddine.bouhaddi@femto-st.fr}

\affiliation[label1]{organization={University of Franche-Comté, CNRS, FEMTO-ST Institute, Department of Applied Mechanics},
            addressline={26 Rue de l'Épitaphe}, 
            city={Besançon},
            postcode={25000}, 
            state={Franche-Comté},
            country={France}}

\begin{abstract}
In this paper, the vibration energy localization in coupled nonlinear oscillators is investigated, based on the creation of standing solitons. The main objective is to establish a design methodology  for mechanical lattices using the Nonlinear Schrödinger Equation (NLSE) as a guide strategy, even in the presence of damping. A three-dimensional diagram is used to illustrate stable parameter regions for damped stationary solitons. Moreover, an analysis of the influence of the number of oscillators in the system, and a numerical investigation regarding the stability of solitonic behavior is done. Through numerical analyses, it is observed that the developed algorithm not only has the capability to locate the highest amplitudes in the chain of oscillators, but also to control the intensity at which these amplitudes are located according to design requirements.The outcomes of the proposed methodology elucidate the impact that the coupling stiffness has on the stabilization of the NLSE, as well as the influence of the number of oscillators on the continuity hypothesis. The developed algorithm holds potential for practical applications in mechanical engineering since the NLSE is used as a design line rather than as a consequence of the phenomenon description. 
\end{abstract}

\begin{highlights}
\item An analysis regarding the minimal number of oscillators for the continuity hypothesis
\item A description of the impact of the coupling stiffness on solitonic behavior
\item An investigation of localized modes from an initial motionless lattices
\end{highlights}

\end{frontmatter}

\section{Introduction}\label{sec1}

The study of coupled nonlinear oscillators has been extensively documented in the literature \cite{strogatz1992coupled,morgante2002standing,chandrasekar2012class,sone2022topological, bukhari2023breather, lenci2022exact, polczynski2021nonlinear}. One characteristic that sets apart linear chains from nonlinear ones is the capacity of these systems to reproduce Intrinsic Localized Modes (ILM) of vibration, even when considering periodicity, whereas, in linear systems, this phenomenon is exclusive with the inclusion of impurities \cite{zergoune2019energy}. In this context, research focusing on localizing vibration energy in these systems carries promise for a range of applications, including innovative sensing technologies  \cite{pandit2019utilizing,grenat2022mass,spletzer2008highly} and dynamic analysis in structures  \cite{manav2018mode}, for instance.

In parallel with the studies on energy localization in nonlinear systems \cite{wang2020irreversible}, works in the area of solitons in mechanical structures has growth in recent years \cite{kenig2009intrinsic,grolet2016travelling,fontanela2018dark,savadkoohi2017nonlinear,jallouli2017stabilization,fontanela2019dissipative,adile2021dynamics}. The convergence of these two areas of study presents a novel strategy for energy localization: the replication of stationary solitons in chains of nonlinear oscillators. The theoretical reproduction of solitons in such systems has been reported in various works, encompassing micromechanical systems \cite{kenig2009intrinsic}, coupled pendula \cite{jallouli2017stabilization}, rotors \cite{fontanela2019dissipative}, and linear architecture oscillators \cite{adile2021dynamics}.

One of the solitonic equations of particular interest for mechanical lattices/chains is the Nonlinear Schrödinger Equation (NLSE). This equation is capable of describing a variety of nonlinear phenomena beyond mechanical context. Recent examples have reported the use of NLSE in biology \cite{boopathy2022nonlinear} and photonics \cite{hernandez2022soliton}, for example. From this perspective, the search for solutions of NLSE is of special interest not only for mechanical engineering but also for the study of nonlinear phenomena in general. However, unlike other research areas (such as optics \cite{shehzad2023multi}), the presence of damping usually cannot be neglected, which significantly increases the difficulty in finding solutions for NLSE.

Studies investigating the feasibility of reproducing mechanical standing solitons in the presence of damping can generally be categorized based on the type of excitation applied to the studied systems \cite{BARBOSA2023110879} . Examples include parametric excitation \cite{kenig2009intrinsic}, external excitation \cite{barashenkov1996existence}, or a combination of both \cite{bitar2017collective}. These investigations provide theoretical evidence for the viability of these phenomena, however, to the best of our knowledge, there is a lack of experimental validations concerning stationary solitons in chains of macro oscillators, particularly when subjected to external excitation, as no known analytical solutions have been reported for the NLSE derived of it \cite{barashenkov1996existence}.

Given the few number of studies that explore the relationship between the stability of stationary solitons and the physical parameters of externally damped resonators, the objective of this research is to establish a design methodology for chains of nonlinear oscillators under external excitation in order to create stationary solitons, which has immediate parallel with localized vibration modes. The goal of the article is to propose a design algorithm employing the externally driven and damped NLSE as the central guiding principle, which, as far as we know, has not yet been reported.

A significant number of research concerning energy localization in nonlinear lattices on a solitonic background either disregards high damping values \cite{thota2015harnessing,kroon2010appearance,sato2006colloquium,feng2006regularized,johansson2002standing,peyrard1998pathway,maluckov2008solitons, grolet2016travelling}  or either relies other forms of excitation \cite{jallouli2017stabilization, gzal2023analysis} or either do not engage in a discussion of how the NLSE can be used in the design of the physical parameters of the system \cite{sato2008visualizing,sato2011experimental,khomeriki2001pattern,rosanov2012knotted,khomeriki2002excitation,hennig2023dissipative,yu2011resonant,moleron2014solitary,nistazakis2002targeted,tchameu2014mobility, fontanela2019dissipative}; which constitutes a distinction of this article from what is already documented. As a result of the proposed methodology the following novelties are presented: (I) an analysis of the influence of the number of oscillators in the periodic structure; (II) an investigation into the impact of the coupling stiffness on the lattice and (III) a numerical investigation of localized modes from an initial motionless configuration of the lattice. A construction of a three-dimensional diagram of stable parameter regions for externally driven damped stationary solitons (based on the two-dimensional case \cite{barashenkov1996existence}) is also done, in order to clarify the effects of physical parameters on solitonic stability. As presented throughout the development, there are optimal choices of design parameters when high damping values are taken into account, which adds another layer of originality to the study.

So far, a series of studies devoted to the conditions for the existence of ILM is reported in the literature \cite{thakur2007driven, ikeda2013intrinsic, balachandran2015response, ikeda2015intrinsic, perkins2019restricted}. However, few papers propose to manipulate such phenomena. In \cite{thota2015harnessing}, the authors analyzed how possible impurities could move the ILM along the chain. Another manipulation strategy was introduced by \cite{lee2021new}, where the authors presented a structure that can change the properties of the system (magnetic springs) in real time. In this context, the design methodology developed in this study adds to the issue of ILM control by enabling the designer to choose the intensity at which the vibration localization can occur since, as is numerically demonstrated, the allocation of oscillators in regions of higher amplitudes is also a design variable.

The rest of the paper is organized as follows: Section 2 shows the mathematical description of the solitonic equation that describes the oscillation amplitude of the chain, where a nondimensionalization of the physical variables is performed. Section 3 presents the relationship between physical parameters that enable the reproduction of the NLSE. Section 4 discusses the development of a design algorithm, while Section 5 the numerical validation of the methodology is evaluated. Finally, Section 6 concludes the paper and provides insights into possible future directions.

\section{NLSE in externally driven damped oscillator chains}\label{sec2}

\begin{figure}[ht!]
  \centering
  \begin{tabular}{@{}c@{} }
    \includegraphics[scale=.3]{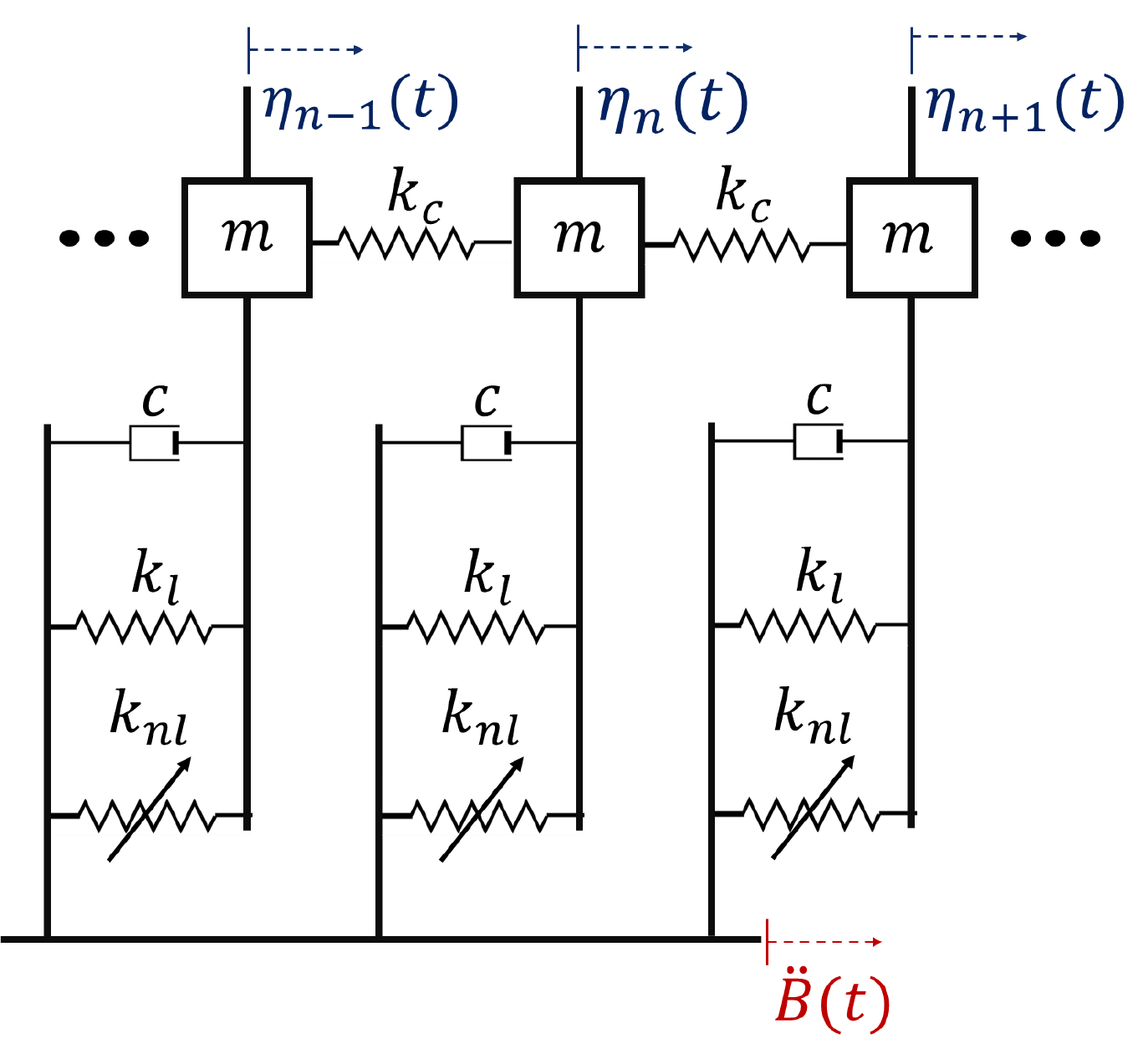} \\[\abovecaptionskip]
    \small (a) 
  \end{tabular}
 
  \begin{tabular}{@{}c@{}}
    \includegraphics[scale=.3]{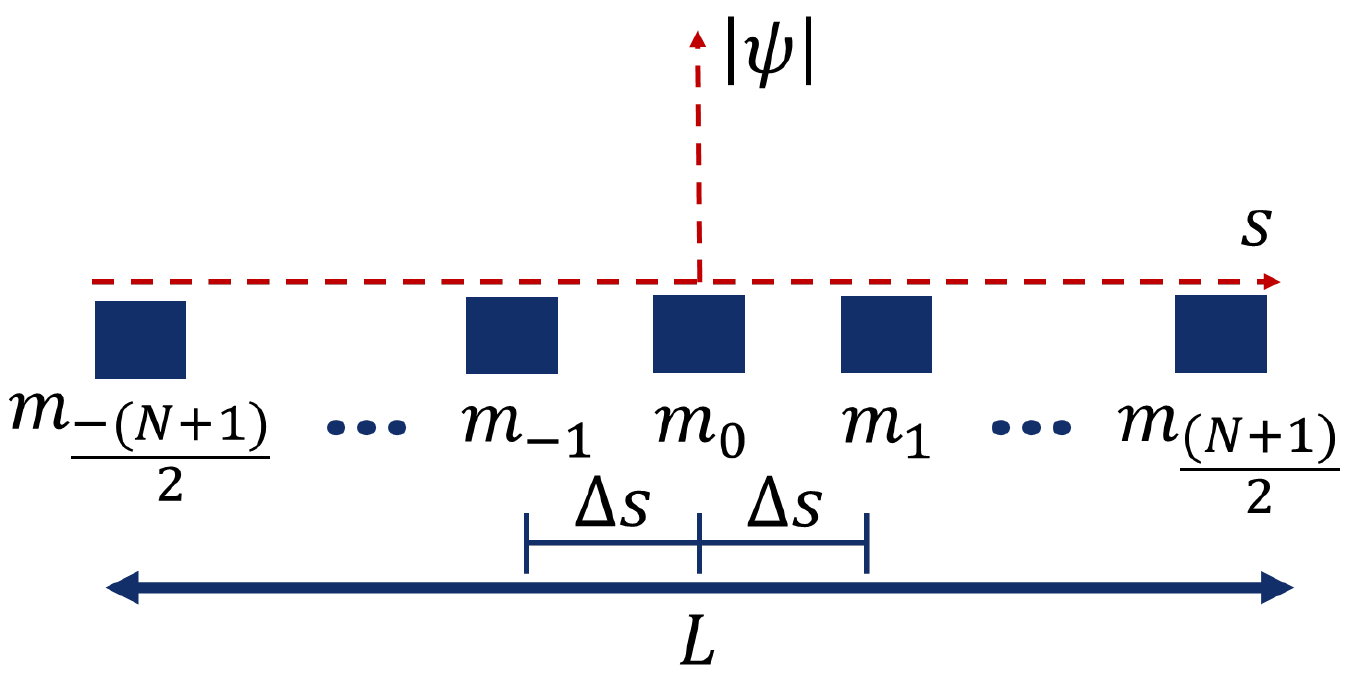} \\[\abovecaptionskip]
    \small (b) 
  \end{tabular}

  \caption{Equivalent mechanical system describing coupled Duffing Oscillators (a), and their arrangement along the chain (b)}
  \label{Fig1}
\end{figure}

The system depicted in Fig. \ref{Fig1}a represents a chain of linearly coupled Duffing oscillators. This system aims to be general, not limited to a specific physical architecture. The index $n$ denotes the displacement $\eta$ of the  n-th oscillator under a base acceleration $\ddot{B}$, which, like $\eta$, is a function of time $t$. The oscillators are connected to each other by a linear stiffness $k_c$ and to the base by a linear stiffness $k_l$, a nonlinear stiffness $k_{nl}$, and a damping coefficient $c$. The excitation of the system occurs through the acceleration imposed on the base. Mathematically, the description of the system is analogous when an external force is considered. With the configuration shown in Fig. \ref{Fig1}a, the individual cells will be subjected to the same acceleration.

\subsection{Proposed solution}
\label{sec:2.1}
 
The equation that describes the system can be obtained by analyzing the relationship between the forces acting on each oscillator, leading to the following expression:

 \begin{equation}
    \begin{array}{c}
        \\ \ddot{\eta}_n+2 \zeta \omega_0 \dot{\eta}_n+\omega_0^2 \eta_n -\frac{k_{n l}}{m} \eta_n^3
         +\frac{k_c}{m}\left(2 \eta_n-\eta_{n+1}-\eta_{n-1}\right)
        \\= -\ddot{B} = -B_A cos(\omega t) = -B_A (0.5 e^{i \omega t}+0.5 e^{-i \omega t} )
    \end{array}
    \label{eq1} 
\end{equation}where
 
 \begin{equation}
        \begin{array}{c c c c}
        \omega_{0}^2=\frac{k_l}{m}, ~&~ 2 \zeta \omega_{0} = \frac{c}{m}, ~&~ \eta_{n}(t) =\eta_{n}, ~&~ k_{nl} \rightarrow -k_{nl}.
    \end{array}
    \label{eq2}
\end{equation}The nonlinear stiffness, along with the other physical parameters, is a design variable that needs to be determined. As demonstrated in this section, the choice of a negative nonlinear stiffness, for a positive coupling stiffness, should align with the sign of the external excitation. Additionally, a stationary boundary condition will be considered: $\eta_0 = \eta_{N+1} = 0$.

The solution of Equation (\ref{eq1}) can be obtained using the method of multiple scales, where, through a perturbation parameter $\epsilon << 1$, the dynamical equation can be rewritten as:
\begin{equation}
    \begin{array}{c}
        \\ \ddot{\eta}_n+2 \epsilon \zeta_\epsilon \omega_0 \dot{\eta}_n+\omega_0^2 \eta_n -\frac{\epsilon k_{n l \epsilon}}{m} \eta_n^3
         +\frac{\epsilon k_{c\epsilon}}{m}\left(2 \eta_n-\eta_{n+1}-\eta_{n-1}\right)
        \\=  -\epsilon B_{A\epsilon} (0.5 e^{i \omega t}+0.5 e^{-i \omega t} ),

    \end{array}
    \label{eq3} 
\end{equation}where the physical parameters, when indexed with $\epsilon$, are rescaled as functions of $\epsilon$.

According to \cite{nayfeh2008nonlinear} the format of the solution of Equation (\ref{eq3}) can be derived using a function whose amplitude varies with time:

\begin{equation}
\eta_{n} = -(\psi_n\left(T\right) e^{-i t \omega }+e^{i t \omega } \overline{\psi}_n\left(T\right)) + O(\epsilon) + O(\epsilon^2)...,
    \label{eq4}
\end{equation}where $O(\epsilon) + O(\epsilon^2)...$ incorporates the terms of the solution proportional to $\epsilon$ or its powers, and the notation $\overline{\psi}_n$ indicates the complex conjugate of $\psi_n$. A new time scale $T = \epsilon t$ is introduced, which controls the temporal behavior of the amplitude $\psi_n$. Equation (\ref{eq4}) has a negative sign accompanying the first term of the solution. Since $\psi_n$ is a function that needs to be deducted, the choice of the sign accompanying it is arbitrary. This choice is convenient for extracting the NLSE, as is detailed in the following section.

\subsection{Continuity Hypothesis}

Fig. \ref{Fig1}b illustrates the positioning of the resonators along the $s$ axis. According to this configuration, for a spacing $\Delta s$, the length $L$ of the structure is related to the number $N$ of oscillators by the expression:

\begin{equation}
\Delta s=\frac{L}{N-1}.
\label{eq5}
\end{equation}

Although Equation (\ref{eq5}) is simple from an algebraic point of view, it suggests that, given a value of $L$, for a sufficiently large number of oscillators, the system shown in Fig. \ref{Fig1}a can be approximated as a continuous domain, where the oscillation amplitudes can be expressed as functions of the oscillator positions $s_n$: 

\begin{equation}
\psi_n(T) \approx \psi(n \Delta s, T)=\psi\left(s_n, T\right) = \psi.
\label{eq6}
\end{equation}The continuity hypothesis assumes that the solution $\psi_n$ varies smoothly and continuously with respect to the spatial coordinate $n$ in the lattice.

Another necessary consideration  for the extraction of the NLSE relates to the amplitudes of adjacent oscillators. By using the Taylor expansion, under the assumption of continuity, the following approximations are made:

\begin{equation}
\psi_{n \pm 1} \approx \psi \pm \Delta s \frac{\partial}{\partial s} \psi+0.5 \Delta s^2 \frac{\partial^2}{\partial s^2} \psi.
\label{eq7}
\end{equation}

\subsection{Obtaining the NLSE}
\label{sec:2.2}

Considering an external excitation frequency $\omega = \omega_0(1 - \epsilon)$, substituting Equation (\ref{eq4}) into Equation (\ref{eq3}) and applying the continuity hypothesis, for approximations up to order $\epsilon^2$, we obtain:

\begin{equation}
\epsilon (F + \overline{F} -\frac{k_{n l \epsilon}}{m}\psi^3 e^{3 i\omega_0 ( T - t)}  ) + \epsilon O(\epsilon)   = 0
\label{eq8} 
\end{equation} where

 \begin{equation}
    \begin{array}{c}
        \\ F = 
        \\  \left(0.5 B_{A\epsilon}    +2 i \zeta_\epsilon \omega_0^2  \psi   +  \frac{3 k_{n l \epsilon}  \psi^2     \overline{ \psi }}{m} -2 \omega_0^2  \psi) \right) e^{i \omega_0(T - t )}
        \\  \left(+ \frac{k_{c e} \Delta s^2}{m} \frac{\partial^2}{\partial s^2}  \psi + 2 i \omega_0    \frac{\partial}{\partial T}  \psi\right)e^{i \omega_0(T - t )}.
    \end{array}
    \label{eq9} 
\end{equation}

By using Equation (\ref{eq8}), the solution of Equation (\ref{eq3}) can be developed by controlling the secular terms proportional to $e^{\pm i \omega_0(T - t)}$. This control is accomplished by setting $F$ to zero, since
 \begin{equation}
    F = 0 \Longleftrightarrow \overline{F} = 0.
    \label{eq10} 
\end{equation}Dividing Equation (\ref{eq9}) by $2 \omega_0^2$, we can obtain the NLSE in the following form:

  \begin{equation}
        \begin{array}{c}
        \\ \frac{0.25 m}{ k_l} B_{A\epsilon}+i\zeta_\epsilon  \psi +\frac{3k_{n l \epsilon}}{2  k_l} \psi^2  \overline{\psi} - \psi  + \frac{k_{c e} \Delta s^2}{k_l} \frac{\partial^2}{\partial s^2}\psi   
        \\ =
        \\-\frac{i}{ \omega_0}\frac{\partial}{\partial T}  \psi.
    \end{array}
    \label{eq11}
\end{equation}

\subsection{Adimensionalization of the NLSE}

Equation (\ref{eq11}), although describing a nonlinear wave, does not necessarily guarantee the existence of stationary solutions ($\frac{\partial}{\partial T} \psi = 0$) for all ranges of physical parameters. To assess this possibility, the following dimensionless variables and relations are added:

 \begin{equation}
        \begin{array}{c c c}

        \\ \hat{\psi}=\sqrt{\frac{3 k_{n l \epsilon}}{4 k_l}}\psi, &  \hat{T}  = \frac{T}{\omega_0}, &\hat{s}=\frac{s}{\delta}, 
        \\ k_{c e}=\frac{2 \delta^2 }{{\Delta s}^2 }k_l , &h_1 = 0.125 \sqrt{3} m B_{A\epsilon}, &h_2 = \sqrt{\frac{k_{n l \epsilon}}{k_{l}^3}},
        
            \end{array} 
    \label{eq12}
\end{equation} where $\delta$, as well as the physical parameters, must be determined by the designer. Relating the expressions in (\ref{eq12}) to Equation (\ref{eq11}), the adimensionalized NLSE is derived as follows:

\begin{equation}
 h_1h_2+\left(2|\hat{\psi}|^2+i \zeta_\epsilon-1\right) \hat{\psi}+\frac{\partial^2 \hat{\psi}}{\partial \hat{s}^2} +i \frac{\partial \hat{\psi}}{\partial \hat{T}} = 0.
 \label{eq13}
\end{equation}Equation (\ref{eq13}) establishes a balance between the damping factor $\zeta_\epsilon$, the externally induced excitation $h_1$, and the nonlinear factor $h_2$. This equation represents a specific scenario within mechanical systems, which do not account for excitation originating from parametric formats or nonlinear damping forms \cite{kenig2009intrinsic}. It should be noted that Equation (\ref{eq13}) is widely reported in the literature without the discrimination of the product $h_1h_2$; however, from an engineering perspective, the influence of the physical parameters should be differentiated: both $h_1$ and $h_2$ work as a external excitation, but have distinct physical origins.

\section{Relationship between Physical Properties and Solitons}
\label{sec:3}

\subsection{Three-dimensional diagram of existence and stability}
\label{sec:3.1}

The work \cite{barashenkov1996existence} analyzed the viability of stationary solutions of Equation (\ref{eq13}). According to the authors, increasing the damping restricts the range of pairs $h_1h_2\times\zeta_\epsilon$ that allow stable solutions. A fundamental difference between this study and the work \cite{barashenkov1996existence} lies in the extrapolation of the product $h_1h_2$ and the discrimination of their physical natures. Previously, the results were obtained considering a single variable $h = h_1h_2$. From an experimental point of view, the detailed characterization of not only the external excitation $B_{A\epsilon}$ but also the nonlinearity $k_{n l \epsilon}$ must be taken into account in the design of the structure, which is why the construction of variable $h_1$, as a function of the base acceleration, and variable $h_2$, as a function of non-linear stiffness, must be done.

Based on the results of \cite{barashenkov1996existence}, the surfaces that delimit values of $h_1h_2\times\zeta_\epsilon$ for stationary solitons described by Equation (\ref{eq13}) are given by the expressions summarized in the Table \ref{tab1}.

\begin{table*}[ht]
\caption{Surfaces used in the construction of the stability diagram shown in Figure \ref{Fig2}}
\label{tab1}
\begin{tabular}{p{0.25\linewidth}c}
\hline 
Surface & Equation  \\
\hline 
black \\$0\leq \gamma \leq 0.66$ & $h_1h_2 = \frac{2\gamma}{\pi} $ \\\\
purple \\$0\leq \gamma \leq 0.3$ & $h_1h_2 = (1.8728 \gamma^3+1.6822\gamma^2+0.0997\gamma+0.0777) $\\\\
blue \\$0.3\leq \gamma \leq 0.5$ & $h_1h_2 = \left(\frac{1}{3}\left(\frac{1}{3}\left(-\gamma^2+\frac{1}{3}\right)^3\right)^{0.5}+\frac{1/3}{3}\left(\gamma^2+\frac{1}{9}\right)\right)^{0.5}$\\\\                          
yellow \\$0.5\leq \gamma \leq 0.66$ & $h_1h_2 = \left(\gamma^3-\gamma^2+\frac{\gamma}{2}\right)^{0.5}$\\
\hline 
\end{tabular}
\end{table*}

 \begin{figure*}[ht]
  \centering
  \begin{tabular}{@{}c@{}c@{}}
    \includegraphics[scale=.3]{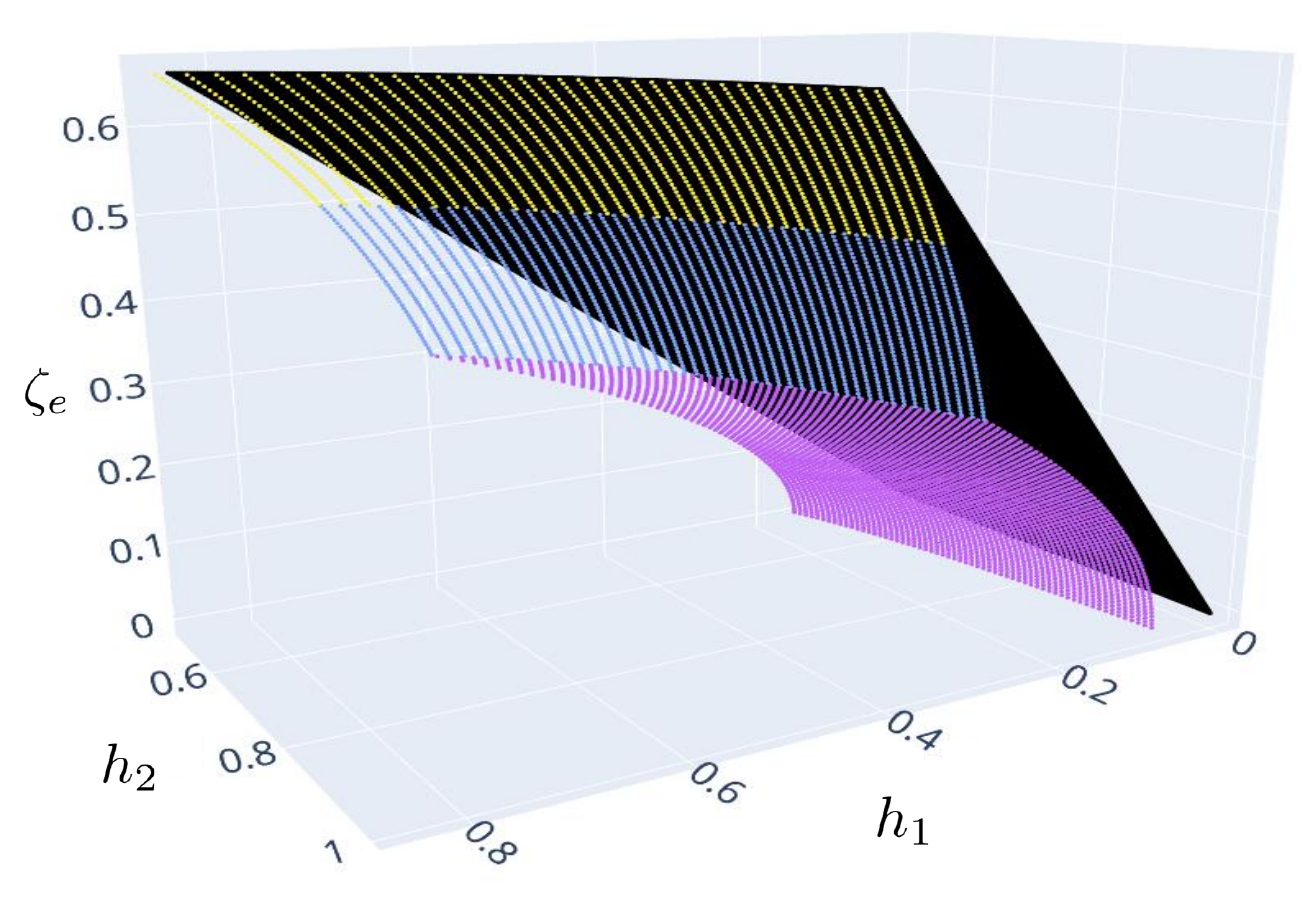} & \includegraphics[scale=.3]{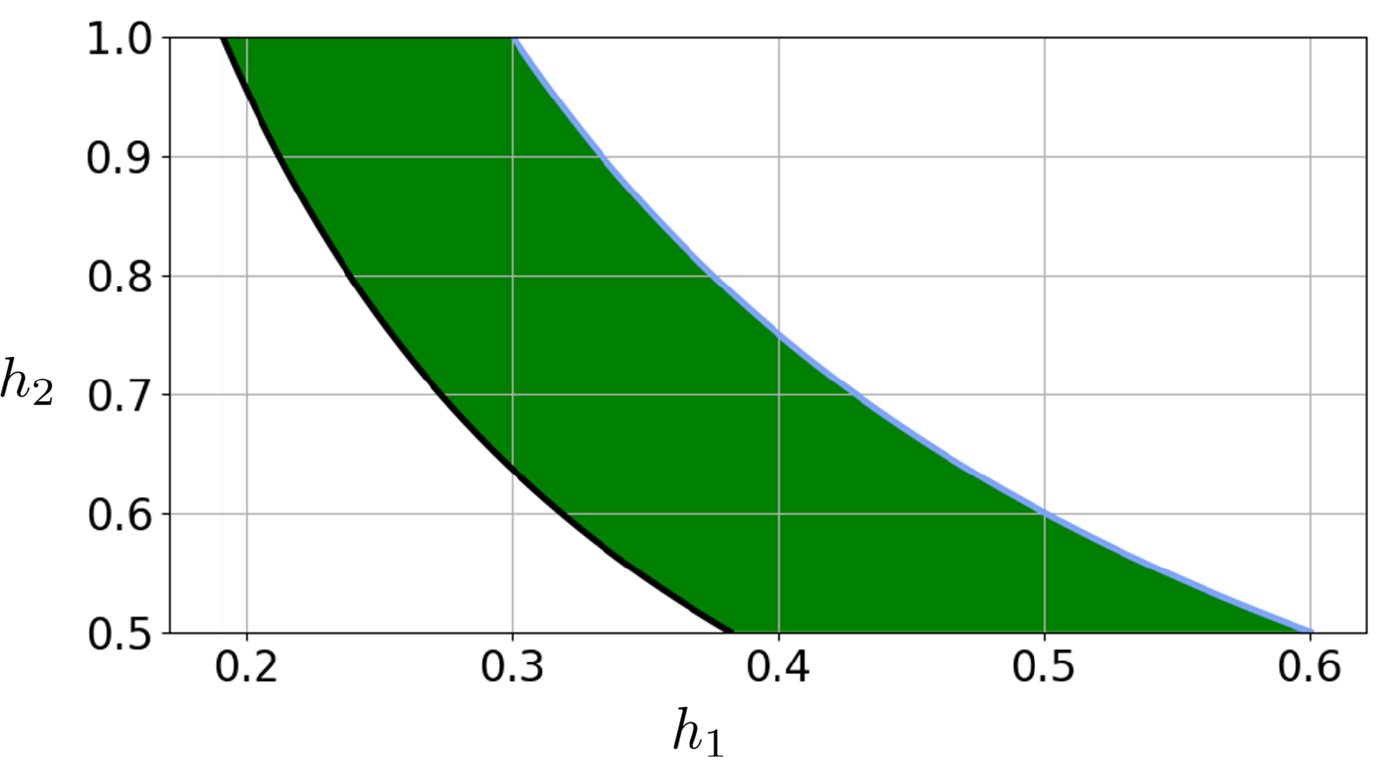} \\
    \small (a) & \small (b)
  \end{tabular}
  
  \vspace{\floatsep}
  
  \begin{tabular}{@{}c@{}}
    \includegraphics[scale=.4]{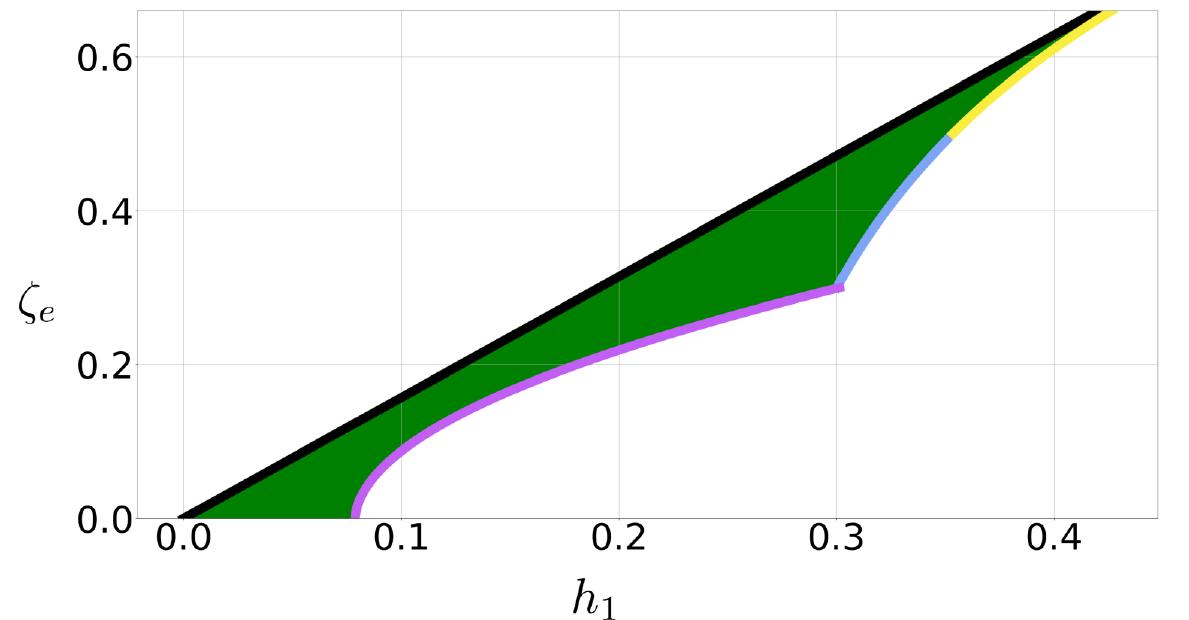} \\
    \small (c)
  \end{tabular}
  
  \caption{Three-dimensional parameter diagram for stable solutions of Equation (\ref{eq13}) (a), particularization of the diagram for $\gamma = 0.3$, and particularization of the diagram for $h2 = 1$}
  \label{Fig2}
\end{figure*}

In agreement with \cite{barashenkov1996existence}, Surface 1, although is a good approximation for the solitonic existence boundary \cite{kaup1978solitons}, is not analytically derived. For values of $\zeta_\epsilon < 0.5$, its accuracy in describing the boundary was indistinguishable from numerical analyses. Surface 2 describes Hopf bifurcation, which forms a boundary between solutions with temporal stability. Its approximation was obtained using the least squares method based on the numerical results presented in \cite{barashenkov1996existence}.

Fig. \ref{Fig2}a displays the created three-dimensional stability graph using the mentioned expressions, which will be used in the design of the oscillator chain. By selecting the value of $\zeta_\epsilon$ on the diagram, the surfaces define an area (see Fig. \ref{Fig2}b) in which the pairs $h_1\times h_2$ ensure numerical convergence through the Newton's method \cite{barashenkov1996existence,jallouli2017stabilization,bitar2017collective}. As can be observed, increasing $\zeta_\epsilon$ reduces the possible area of stationary solutions.

The diagram shown in Fig. \ref{Fig2}a is a generalization of the particular two-dimensional case (Fig. \ref{Fig2}c). When the value of $h_2$ is fixed ($h_2 = 1$, for instance), the graph deduced by \cite{barashenkov1996existence} is obtained, keeping in perspective the distinction between the orientation of the axes and the omission of regions beyond Hopf bifurcation.

\subsection{Physical properties along the lattice}
\label{sec:32}

The normalization of Equation (\ref{eq11}) allows the methodology to generalize the analysis beyond specific values of mass, stiffness, or resonator architecture. Rescaling relations (\ref{eq12}) by considering the perturbation factor $\epsilon$, the physical parameters can be related to the dimensionless ones by the following equations:

\begin{equation}
 \zeta  =  \epsilon \zeta_\epsilon,
 \label{eq14}
\end{equation}

\begin{equation}
k_{c}=\epsilon\frac{2 \delta^2}{\Delta s^2} k_l,
\label{eq15}
\end{equation}

\begin{equation}
B_A = \epsilon \frac{h_1 }{0.125 \sqrt{3} m},
\label{eq16}
\end{equation}

\begin{equation}
k_{nl} = \epsilon k_l^3 h_2^2,
\label{eq17}
\end{equation}where, as a condition for solitonic existence, $(\zeta_\epsilon, h_1, h_2)$ are contained in the volume delimited by the stability conditions (Fig. \ref{Fig2}a).

The quantity $\zeta_\epsilon$ is bounded by $0 \leq \zeta_\epsilon \leq 0.66$, which implies limits on the physical damping ratio $0 \leq \zeta \leq \epsilon 0.66$. This relation has a practical implication for a potential oscillator design: high damping values necessarily correspond to low values of $\epsilon$. Considering that the coupling stiffness, the base acceleration amplitude, and the nonlinear stiffness are directly proportional to $\epsilon$ (Equations (\ref{eq15}) - (\ref{eq17})), the choice of a feasible $\epsilon$ (within the possibilities of experimental implementation) should also guide the design methodology.

The parameter $\delta$ was introduced previously as responsible for the adimensionalization of the space (Equations (\ref{eq12})). The choice of $\delta$ is directly linked to the parallel between the length $L$ of the lattice and the domain $\hat{s}$ in Equation (\ref{eq13}). 

\begin{figure}[ht]
\centering
\includegraphics[scale=.35]{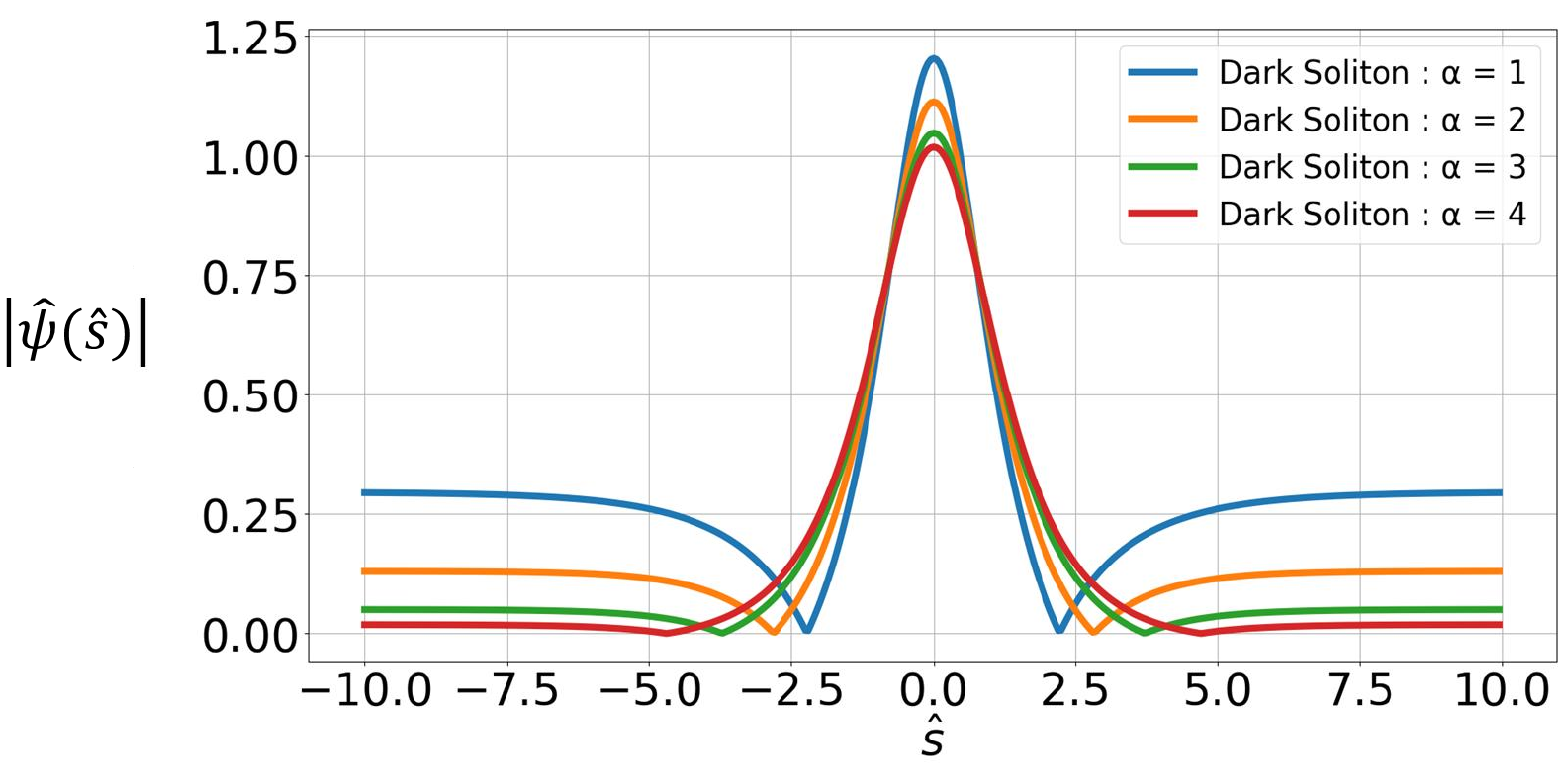}
\caption{Module of  $\hat{\psi}(\hat{s})$ for various values of $\alpha$ applied to Equation (\ref{eq18})}
\label{Fig3}     
\end{figure}

To illustrate the role of $\delta$, let us consider the non-damped ($\zeta_\epsilon = 0$) stationary dark soliton solution of Equation (\ref{eq13}) proposed by \cite{barashenkov1996existence} \footnote{``Dark'' notation employed in contrast to the other potential solution, unstable in time when damping is introduced, also presented in \cite{barashenkov1996existence}.}:

\begin{equation}
\hat{\psi}(\hat{s})=\hat{\psi}_0\left(1+\frac{2 \sinh ^2 \alpha}{1 - \cosh \alpha \cosh (A \hat{s})}\right),
\label{eq18}
\end{equation} where 
\begin{equation}
        \begin{array}{c c}
        \\ h_1h_2=\frac{\sqrt{2} \cosh ^2 \alpha}{\left(1+2 \cosh ^2 \alpha\right)^{3 / 2}}, ~&~  \hat{\psi}_0=\frac{1}{\sqrt{2\left(1+2 \cosh ^2 \alpha\right)}}, 
        \\ \\ A=2 \hat{\psi}_0 \sinh \alpha.\\
    \end{array}
    \label{eq19}
\end{equation} Figure \ref{Fig3} displays $ |\hat{\psi}(\hat{s})|$ for some values of $\alpha$. As can be observed, the value of the function decays as $\hat{s}$ moves away from the origin, a phenomenon that also occurs for the damped case \cite{barashenkov1996existence, jallouli2017stabilization}. 

Imposing a low value for $\delta$ implies allocating more oscillators outside the domain of high amplitudes ($-5\leq \hat{s}\leq+5$ for the undamped case), or, precisely, accentuating the phenomenon of localized vibrations. From a physical perspective, high values of $\delta$ would impose high values of coupling stiffness (Equation (\ref{eq15})), which would attenuate the differences in oscillator amplitudes since they would be more tightly coupled, which is in line with the conclusion regarding low values of $\delta$.

The refinement of the continuity hypothesis is related to the reduction in the spacing between oscillators (see Fig. \ref{Fig1}b and Equation (\ref{eq5})). From Equation (\ref{eq15}), we can infer that for a given value of $k_c$, lower values of $\Delta s$ allow for the choice of lower values of $\delta$. In conclusion, the localized vibration modes can be more effectively explored as the continuity hypothesis becomes more refined, a previously expected statement.

\section{Design algorithm proposal}
\label{sec:4}

\subsection{Generation of solitons via chains of motionless oscillators}

\begin{figure}[ht]
\centering
\includegraphics[scale=.55]{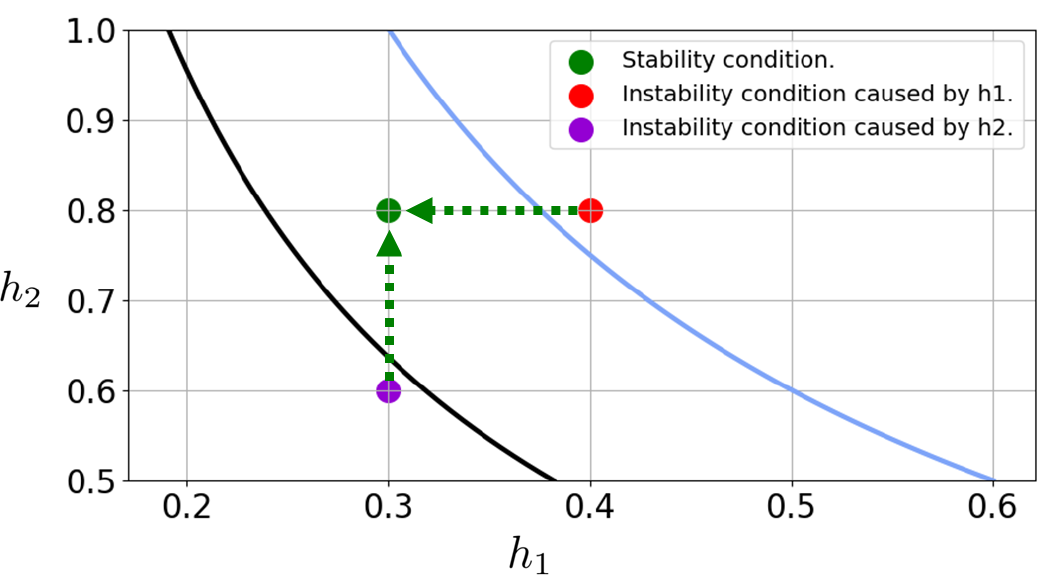}
\caption{Examples of possible initial instability impositions}
\label{Fig4}       
\end{figure}

As pointed out by \cite{kenig2009intrinsic}, the dynamic formation of solitons from a motionless array of resonators is not a straightforward task. It requires taking the system far enough from the basin of attraction of other stable solutions of Equation (\ref{eq13}), particularly flat ones \cite{barashenkov2011travelling}. To overcome this problem, a possible strategy is to impose an unstable initial condition on the system, distant from the stability conditions but closer to the solitonic basin of attraction. This strategy can be implemented both experimentally and numerically. Figure \ref{Fig4} illustrates this approach by proposing an initial condition outside the volume that delimits the possible design parameters. 

Carrying out this procedure through $h_2$, with fixed values of $m$, $B_{A\epsilon}$, and $k_l$, the nonlinear stiffness is increased until $h_2$ determines a point outside the solitonic existence boundary. After a few periods of oscillation, the reverse process occurs, and $h_2$ imposes a point within the stability boundary. The work of \cite{kimura2012experimental} utilized a similar strategy by numerically and experimentally verifying this technique while modulating the nonlinear stiffness of an oscillator chain to generate ILMs.

Another way, similar to nonlinearity modulation, is the excitation modulation, as mathematically, the nonlinear stiffness and the base acceleration have similar effects on the dynamic equation. This claim can be concluded through the following transformation:

\begin{equation}
\hat{\eta}_n = \sqrt{\frac{k_{n l}}{m}}\eta_n.
\label{eq20}
\end{equation} Applying Equation (\ref{eq20}) to (\ref{eq1}), the following expression is derived:

 \begin{equation}
    \begin{array}{c}
        \\ \ddot{\hat{\eta}}_n+2 \zeta \omega_0 \dot{\hat{\eta}}_n+\omega_0^2 \hat{\eta}_n - \hat{\eta}_n^3
         +\frac{k_c}{m}\left(2 \hat{\eta}_n-\hat{\eta}_{n+1}-\hat{\eta}_{n-1}\right)
        \\ = -  \sqrt{\frac{k_{n l}}{m}} B_A cos(\omega t),

    \end{array}
    \label{eq21} 
\end{equation}thus, the correlation between nonlinearity and excitation in the dynamic equation becomes obvious, since they appear side by side in the equation.

\subsection{Design methodology}
\label{sec:42} 

Based on the derived relationships, the development of a methodology for designing the oscillator chain becomes viable. The first step in the design process involves identifying the parameters that describe the dynamic behavior of each oscillator. After so, given an external excitation value, the nonlinear and coupling stiffness can be chosen, enabling the emergence of localized modes. This design strategy is summarized in Fig. \ref{Fig5}.

\tikzset{
  startstop/.style={
    rectangle,
    rounded corners,
    minimum width=3cm,
    minimum height=0.5cm,
    text centered,
    text width=6cm,
    draw=black,
    fill=red!20, 
  },
  arrow/.style={
    ->,
    thick,
    shorten >=2pt,
    shorten <=2pt,
  },
}

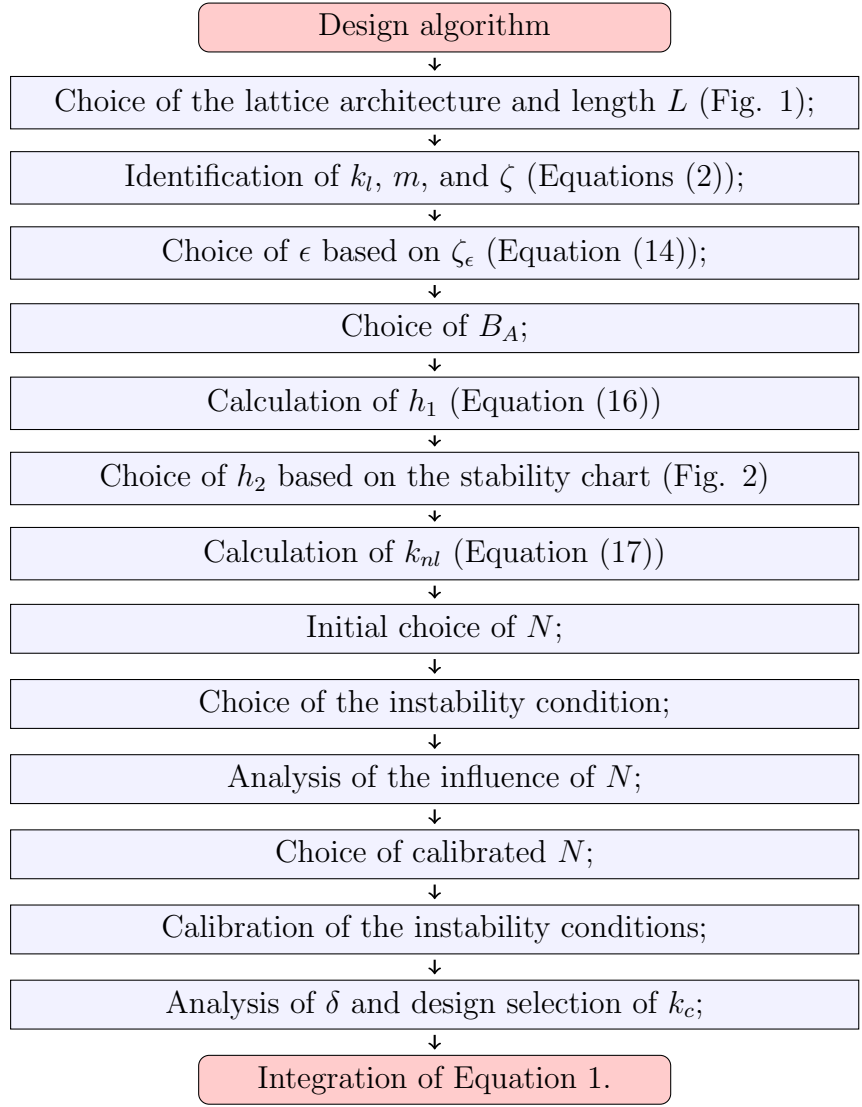
\begin{figure}[!ht]
\centering
\begin{tikzpicture}[node distance=1cm]
\centering

\node (start) [startstop] {Design algorithm};

\tikzstyle{process} = [rectangle, minimum width=10cm, minimum height=0.5cm, text centered, text width=11cm, draw=black, fill=blue!5]
\node (pro1) [process, below of=start] {Choice of the lattice architecture and length $L$ (Fig. \ref{Fig1}); };

\tikzstyle{process} = [rectangle, minimum width=10cm, minimum height=0.5cm, text centered, text width=11cm, draw=black, fill=blue!5]
\node (pro2) [process, below of=pro1, ] {Identification of $k_l$, $m$, and $\zeta$ (Equations (\ref{eq2})); };

\tikzstyle{process} = [rectangle, minimum width=10cm, minimum height=0.5cm, text centered, text width=11cm, draw=black, fill=blue!5]
\node (pro3) [process, below of=pro2, ] {Choice of $\epsilon$ based on $\zeta_\epsilon$ (Equation (\ref{eq14})); };

\tikzstyle{process} = [rectangle, minimum width=10cm, minimum height=0.5cm, text centered, text width=11cm, draw=black, fill=blue!5]
\node (pro4) [process, below of=pro3, ] {Choice of $B_A$;};

\tikzstyle{process} = [rectangle, minimum width=10cm, minimum height=0.5cm, text centered, text width=11cm, draw=black, fill=blue!5]
\node (pro5) [process, below of=pro4,] {Calculation of $h_1$ (Equation (\ref{eq16}))};

\tikzstyle{process} = [rectangle, minimum width=10cm, minimum height=0.5cm, text centered, text width=11cm, draw=black, fill=blue!5]
\node (pro6) [process, below of=pro5, ] {Choice of $h_2$ based on the stability chart (Fig. \ref{Fig2})};

\tikzstyle{process} = [rectangle, minimum width=10cm, minimum height=0.5cm, text centered, text width=11cm, draw=black, fill=blue!5]
\node (pro7) [process, below of=pro6,] {Calculation of $k_{nl}$ (Equation (\ref{eq17}))};

\tikzstyle{process} = [rectangle, minimum width=10cm, minimum height=0.5cm, text centered, text width=11cm, draw=black, fill=blue!5]
\node (pro8) [process, below of=pro7,] {Initial choice of $N$;};

\tikzstyle{process} = [rectangle, minimum width=10cm, minimum height=0.5cm, text centered, text width=11cm, draw=black, fill=blue!5]
\node (pro9) [process, below of=pro8,] {Choice of the instability condition;};

\tikzstyle{process} = [rectangle, minimum width=10cm, minimum height=0.5cm, text centered, text width=11cm, draw=black, fill=blue!5]
\node (pro10) [process, below of=pro9,] {Analysis of the influence of $N$;};
 
\tikzstyle{process} = [rectangle, minimum width=10cm, minimum height=0.5cm, text centered, text width=11cm, draw=black, fill=blue!5]
\node (pro11) [process, below of=pro10,] {Choice of calibrated $N$;};

\tikzstyle{process} = [rectangle, minimum width=10cm, minimum height=0.5cm, text centered, text width=11cm, draw=black, fill=blue!5]
\node (pro12) [process, below of=pro11,] {Calibration of the instability conditions;};

\tikzstyle{process} = [rectangle, minimum width=10cm, minimum height=0.5cm, text centered, text width=11cm, draw=black, fill=blue!5]
\node (pro13) [process, below of=pro12,] {Analysis of $\delta$ and design selection of $k_c$; };

\node (stop) [startstop, below of=pro13] {Integration of Equation \ref{eq1}.};

\draw [arrow] (start) -- (pro1);
\draw [arrow] (pro1) -- (pro2);
\draw [arrow] (pro2) -- (pro3);
\draw [arrow] (pro3) -- (pro4);
\draw [arrow] (pro4) -- (pro5);
\draw [arrow] (pro5) -- (pro6);
\draw [arrow] (pro6) -- (pro7);
\draw [arrow] (pro7) -- (pro8);
\draw [arrow] (pro8) -- (pro9);
\draw [arrow] (pro9) -- (pro10);
\draw [arrow] (pro10) -- (pro11);
\draw [arrow] (pro11) -- (pro12);
\draw [arrow] (pro12) -- (pro13);
\draw [arrow] (pro13) -- (stop);

\end{tikzpicture}

\caption{Design methodology for the active generation of stationary solitons in damped oscillator chains}
\label{Fig5}
\end{figure}

The algorithm described by the 13 steps assumes that the damping is not controlled, while the external excitation is. Other design possibilities could be explored if the external excitation is the uncontrolled parameter. The distinction between these paradigms resides in the active generation (control of $B_A$) or passive generation of solitons in oscillator chains. In this work, the active generation will be considered.

The calibration of the initial conditions is performed through numerical scanning of the time during which the system is subjected to unstable conditions. In this study, the instability is controlled by $h_1$ (Equation (\ref{eq16})). Beyond the manipulation of the external excitation and/or nonlinearity, experimental studies have demonstrated the feasibility of obtaining localized vibration modes through the manipulation of the external excitation frequency \cite{chen2009inducing}. Other procedure for generating solitons would involve initiating the array in a specific non-zero state and driving it outside its known stability boundaries. This approach has been considered in the past in systems with different forms of excitation \cite{wang2001parametrically,barashenkov2003multistable,maniadis2006mechanism, thakur2007driven}.

The methodology for calibrating the number of oscillators is built upon a numerical search for an odd number of oscillators that allows the emergence of the highest number of solitonic behaviors within a range of $\delta$. Choosing an odd value for $N$ ensures that a single oscillator exhibits maximum oscillation amplitude compared to the others.

\subsection{Possible choices of stable parameters}
\label{sec:43}

The stability regions depicted in Fig. \ref{Fig2} have no commitment with the robustness of the parameters concerning potential uncertainties in the physical variables, which inevitably need to be considered in experimental evaluations.

\begin{figure*}[t]
\centering
\includegraphics[scale=.52]{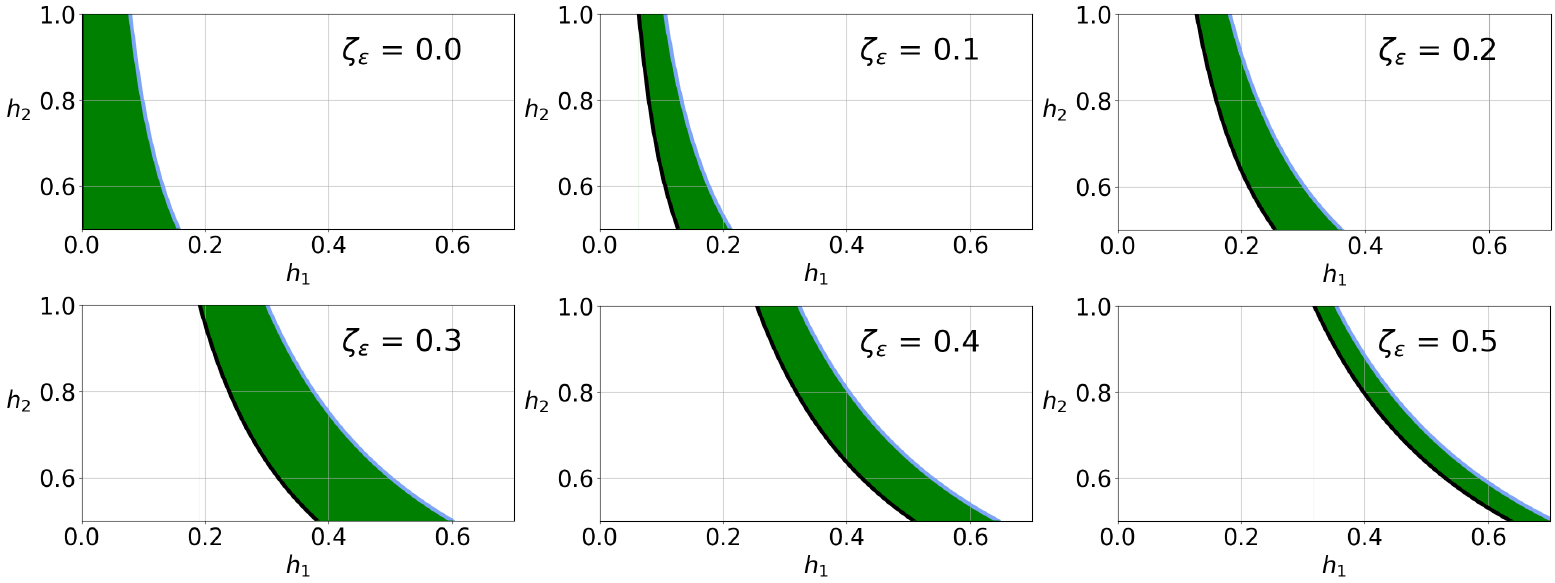}
\caption{Series of stable parameter regions for $h_1$ and $h_2$ according to the variation of $\zeta_\epsilon$}
\label{Fig6}     
\end{figure*}

In order to enhance the robustness of the methodology, a criterion for selecting the area determined by $\zeta_\epsilon$ (see Fig. \ref{Fig2}b) is proposed. Figure \ref{Fig6} provides a depiction of a series of stable parameter regions for different values of $\zeta_\epsilon$. As can be observed, the stable parameter areas vary considerably as a function of $\zeta_\epsilon$, enabling an optimal choice from a robustness perspective.

\begin{figure}[t]
\centering
\includegraphics[scale=.4]{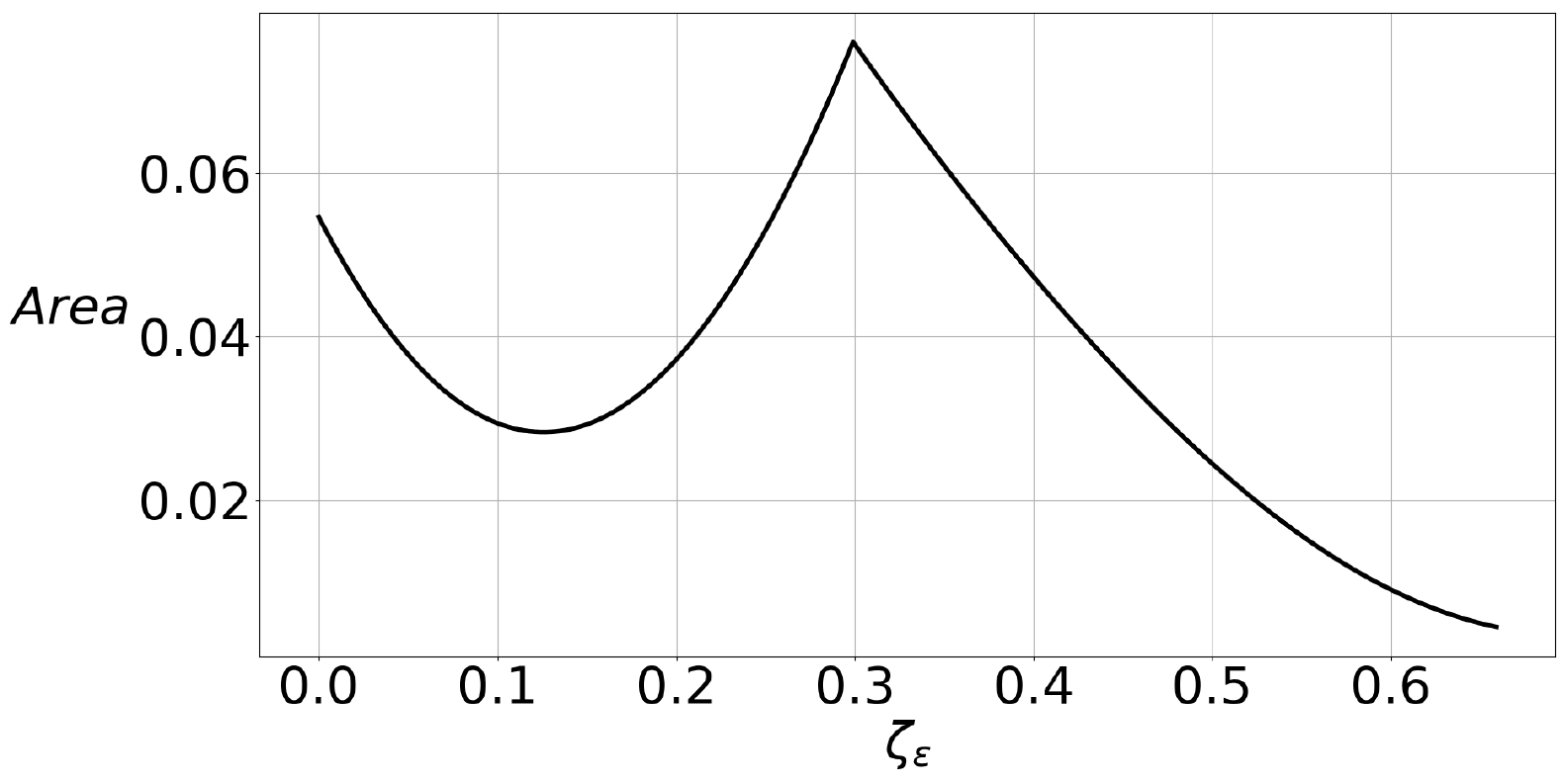}
\caption{Area of stable parameters as a function of $\zeta_\epsilon$}
\label{amortecimento}     
\end{figure}

Figure \ref{amortecimento} shows the relationship between the value of the stability area (Figures \ref{Fig2}b and \ref{Fig6}) as a function of $\zeta_\epsilon$. Despite this graph does not have a direct physical interpretation (since $h_1$ and $h_2$ are determined by Equations \ref{eq16} and \ref{eq17}, respectively), we can observe that the maximization of the stability area occurs around $\zeta_\epsilon = 0.3$. This consideration should be taken into account when choosing an $\epsilon$ value. As an example for, $\zeta = 0.5\%$, choosing $\epsilon = 0.017$ would result in $\zeta_\epsilon \approx 0.3$.

\section{Numerical Validation}

\subsection{Choice of integration method}

The numerical validation of the methodology is carried out through the numerical integration of Equation (\ref{eq1}). This Equation represents a compact notation for a system comprising $N$ nonlinear equations. As previously mentioned, a sufficient large number of oscillators is required for the continuity assumption to be valid. However, in addition to experimental challenges, the computational cost associated with a high number of oscillators can increase significantly. Therefore, it is desirable to have a low number of oscillators that ensures the viability of the methodology.

Not only the number of oscillators but also the dynamic characteristics of Equation (\ref{eq1}) are crucial points in numerical integration. In order to reach convergence, terms that depend on $\eta_n$ must be positive. Consequently, disregarding coupling, the following relationship must be satisfied:
\begin{equation}
\omega_0^2 \eta_n - \frac{k_{n l}}{m} \eta_n^3 \geq 0.
\label{eq22}
\end{equation}
This condition can be achieved by controlling either the value of external excitation or the value of nonlinear stiffness, or both.

An adequate number of oscillators and the respect of the condition imposed by the Inequality (\ref{eq22}) are not enough to guarantee numerical convergence, we must also consider the choice of the numerical method. According to  \cite{wanner1996solving}, stiff equations are problems for which explicit methods are not suitable, as their convergence would require time steps that are not feasible in practical analyses. These types of problems exhibit chaotic behavior, very sensitive to initial conditions. Following this definition, Equation (\ref{eq1}) becomes increasingly stiff as the choices of $h_1$ and $h_2$ come close to the stability boundaries or as the number of oscillators increases.

The work of \cite{wang2017application} implemented the Radau IIA method in the numerical analysis of a flexible multibody system with holonomic constraints, which is a stiff mechanical problem. Previous studies have validated the use of Radau methods applied to similar problems \cite{calvo2000runge,guglielmi2001implementing,ding2009implicit,martin201017th}, indicating the potential application of similar strategies for Equation (\ref{eq1}).

Hence, in order to increase the reliability of numerical integrations, we can select three techniques for the numerical integration: Radau (an implicit Runge-Kutta method of the Radau IIA family with an order of 5), BDF (an implicit multi-step variable-order method based on backward differentiation formulas for derivative approximation), and DOP853 (an explicit Runge-Kutta method of order 8). The first two methods (Radau and BDF) are good candidates for numerical integration of stiff problems due to their implicit nature. The choice of DOP853, despite being explicit, has proven effective in capturing nonlinear dynamics for small enough time steps.

\subsection{Analysis on the number of oscillators}

After the identification of $k_l$ and choosing an appropriate $\epsilon$, $N$ and $\delta$ determine the coupling stiffness (Equation (\ref{eq15})). Mathematically, the coupling stiffness can be understood as a function of the ratio $\left( \frac{\delta}{\Delta s} \right)$. This consideration points to the numerical strategy that will be used to calibrate the number of oscillators in the chain.

As mentioned in Section \ref{sec:32}, the energy localization (related to the oscillation amplitude) will be greater for smaller values of $\delta$. However, the physical consequence of low values of $\delta$ is a decrease in the coupling between oscillators, which, from a physical standpoint, would tend to decouple the resonators. Therefore, the decrease in $\delta$ (intensification of localized vibrational modes) should be accompanied by an increase in the number of oscillators (refinement of the continuity assumption), at the cost of increasing the number of equations to be integrated.

Once the numerical verification of the relationship between $N$ and $\delta$ is complete, it becomes possible to carry out an assessment regarding the duration that the system needs to be exposed to the instability condition. To ensure an appropriate duration for the instability condition, oscillators characterized by lower natural frequencies should undergo it for a longer period, while oscillators with higher natural frequencies should have a shorter exposure. The work \cite{kenig2009intrinsic} utilized a duration of 600 oscillation periods as a benchmark for the instability condition, thereby indicating the approximate timeframe during which the system should be subjected to it.

\subsection{Numerical Example}

In order to implement the steps presented in Fig. \ref{Fig5}, a numerical example is used in accordance with the algorithm's sequence. Following the first seven steps of the design algorithm, possible parameters describing each oscillator are suggested in Table \ref{tab2}. Table \ref{tab3} shows possible parameters taken for the implementation of steps 8 and 9.

\begin{table}
\caption{Parameters used in the numerical example: first 7 steps.}
\label{tab2}        
\begin{tabular}{lll}
\hline\noalign{\smallskip}
Parameter & Value & Unit  \\
\noalign{\smallskip}\hline\noalign{\smallskip}
$L$ & 2 &  $m$ \\
$m$ & 0.1 & $kg$ \\
$\omega_0$ & 314.159 & $rad/s$ \\
$\zeta$ & 0.3\% &dimensionless \\
$\epsilon$ & 0.01 &dimensionless  \\
$\zeta_\epsilon$ & 0.3 &dimensionless  \\
$h_1$  & 2.124 & $N$   \\
$h_2$  & 0.118 & $N^{-1}$   \\
\noalign{\smallskip}\hline
\end{tabular}
\end{table}

\begin{table}
\caption{Parameters used in the numerical example: steps 8 and 9.}
\label{tab3}        
\begin{tabular}{lll}
\hline\noalign{\smallskip}
Parameter & Value & Unit  \\
\noalign{\smallskip}\hline\noalign{\smallskip}
 N & 11  & dimensionless \\
 $h_1$ &  2.973  &  $N$ \\
 $h_2$ &  0.118 & $N^{-1}$   \\
 Unstable regime time & 12  & $s$  \\
\noalign{\smallskip}\hline
\end{tabular}
\end{table}

\subsubsection{Choice of $N$ and analysis of instability conditions}

\begin{figure*}[ht]
  \centering
  \begin{tabular}{@{}c@{}}
    \includegraphics[scale=.5]{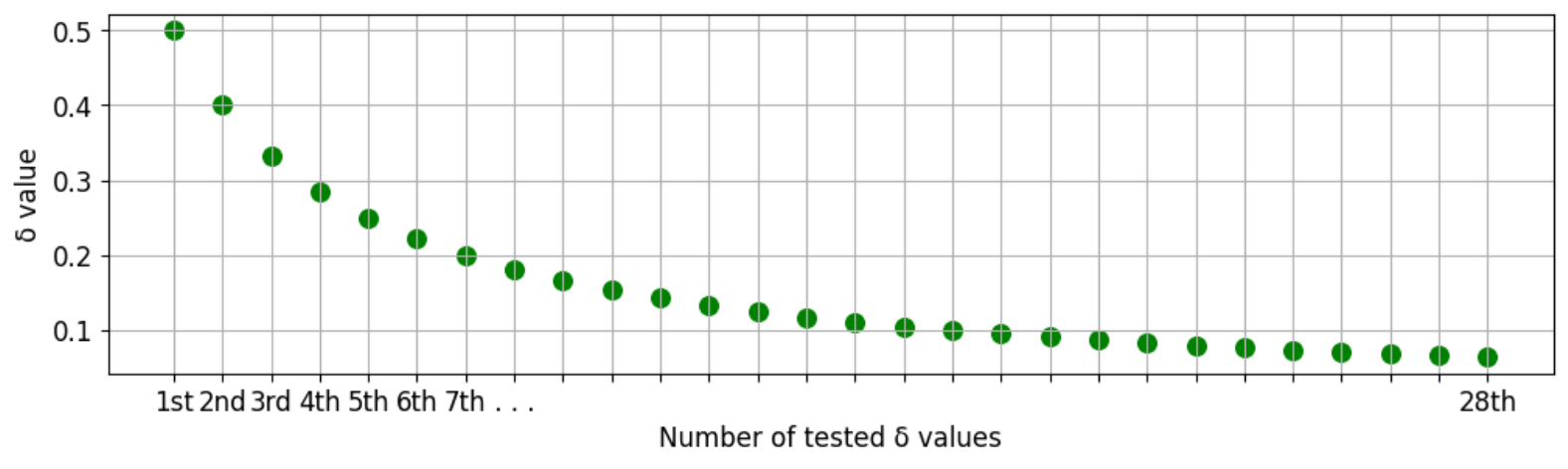} \\[\abovecaptionskip]
    \small (a) 
  \end{tabular}

  \vspace{\floatsep}

  \begin{tabular}{@{}c@{}}
    \includegraphics[scale=.5]{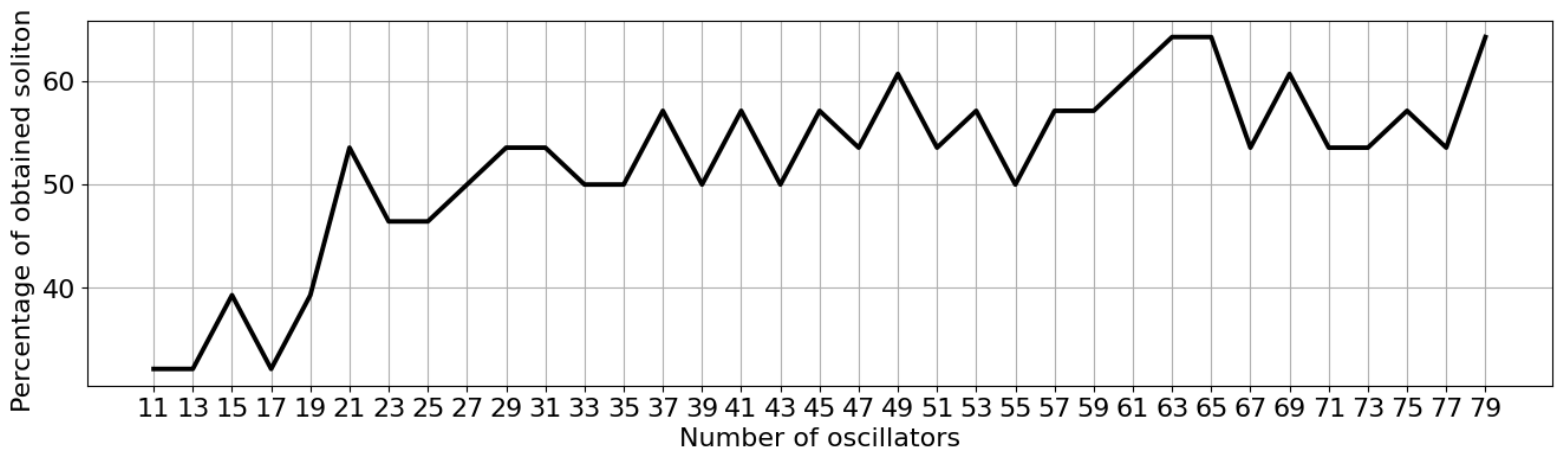} \\[\abovecaptionskip]
    \small (b) 
  \end{tabular} 
  \caption{Values of $\delta$ evaluated in numerical integrations (a), and percentage of solitonic solutions obtained in the range of $\delta$ values as a function of the number of oscillators}
  \label{Fig8}
\end{figure*}

Initially, the system is exposed to an instability condition for a duration of $600$ oscillation periods, as indicated in Table \ref{tab3}. After this time, the system transitions to the solitonic oscillation regime (Table \ref{tab2}), and temporal stability is verified. For a decrease in $\delta$, the procedure is repeated, and after investigating a range of $\delta$ values, the value of $N$ is increased by two units.  Figure \ref{Fig8} presents the results obtained from the numerical integration of Equation (\ref{eq1}) for various values of $N$ within a range of values for $\delta$. We should notice that Figure \ref{Fig8} (b) exhibits fluctuations in the number of solitonic solutions. This is attributed to the selection of the unstable regime submission time.

The integration methods (Radau, BDF, and DOP853) were performed with a time step of $10^{-5}s$ and a total simulation time of $24s$. The investigation examined the existence of solitons for $28$ decreasing values of $\delta$ (Figure \ref{Fig8}a), and the number of stationary solitons was recorded as a function of the number of oscillators (Figure \ref{Fig8}b). As can be observed, starting from $N = 37$, an increase in the number of oscillators does not lead to a significant increase in the number of solitonic solutions for the selected $\delta$ range. Given the good numerical performance for $N = 49$, this value will be chosen as the number of oscillators. 

As expected, increasing the number of oscillators leads to an increase in the number of solitonic dynamics in the oscillator chain. The appearance of such dynamics is strongly influenced by the refinement of the time step used in the integrations. In addition to the time step,  is recommended the selection of multiple integration methods.

Given a fixed value of N and a low value of delta ($\delta < 0.2$), a new numerical scan is performed to determine the temporal interval of instability regime that, after the imposition of stable conditions, leads to ILMs. Taking $N = 49$ and $\delta = 0.1$, it was found that an interval between $9s$ and $12s$ ensures this condition.

\subsubsection{$\delta$ analysis and $k_c$ selection}

\begin{figure*}[ht]
\centering
\includegraphics[scale=.55]{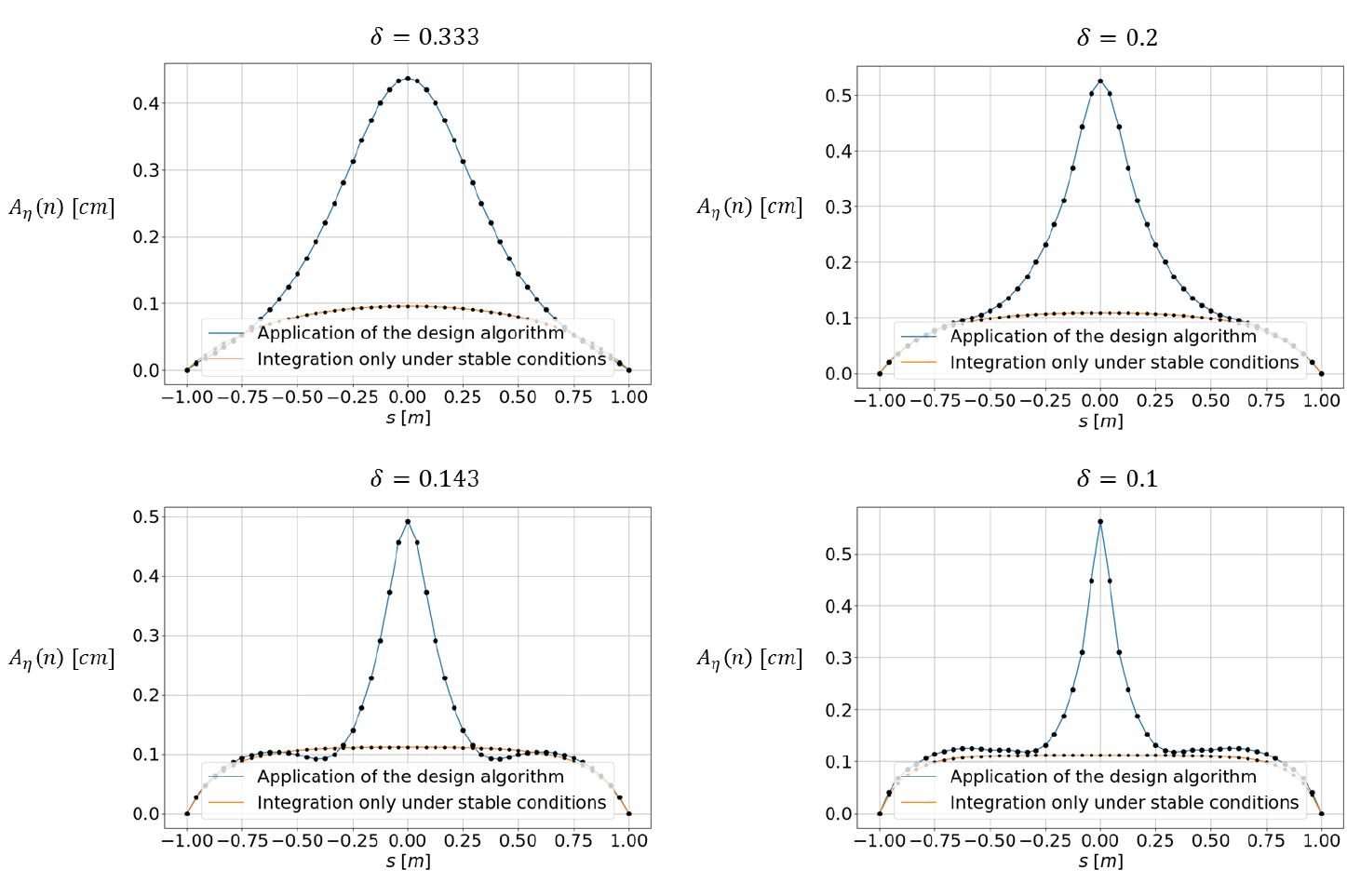}
\caption{Oscillation amplitude $A_{\eta}$ of the N oscillators plotted according to the their position for different values of $\delta$}
\label{Fig9}        
\end{figure*}

The final step of the design algorithm is the selection of the coupling stiffness as a function of $\delta$. Figure \ref{Fig9} shows the maximum oscillation amplitude $A_{\eta}(n)$ of the structure in steady state as a function of the oscillators' positioning $n$ for four values of $\delta$ (amplitude as being understood as twice the value of the maximum $\eta$ value, since the oscillation occurs symmetrically). As predicted, a decrease in $\delta$ allocates fewer oscillators in the region of maximum amplitude (close to the origin in $s = 0$).

Equation (\ref{eq15}) imposes a quadratic relationship between $\delta$ and $k_c$, which restricts practical choices of $\delta$ values from an experimental perspective. For $\delta = 0.2$, $15$ oscillators have vibration amplitudes greater than or equal to $0.2~cm$, whereas, for $\delta = 0.1$, this number decreases to $7$ oscillators using a coupling stiffness four times smaller.

\begin{figure*}
\centering
\includegraphics[scale=0.7]{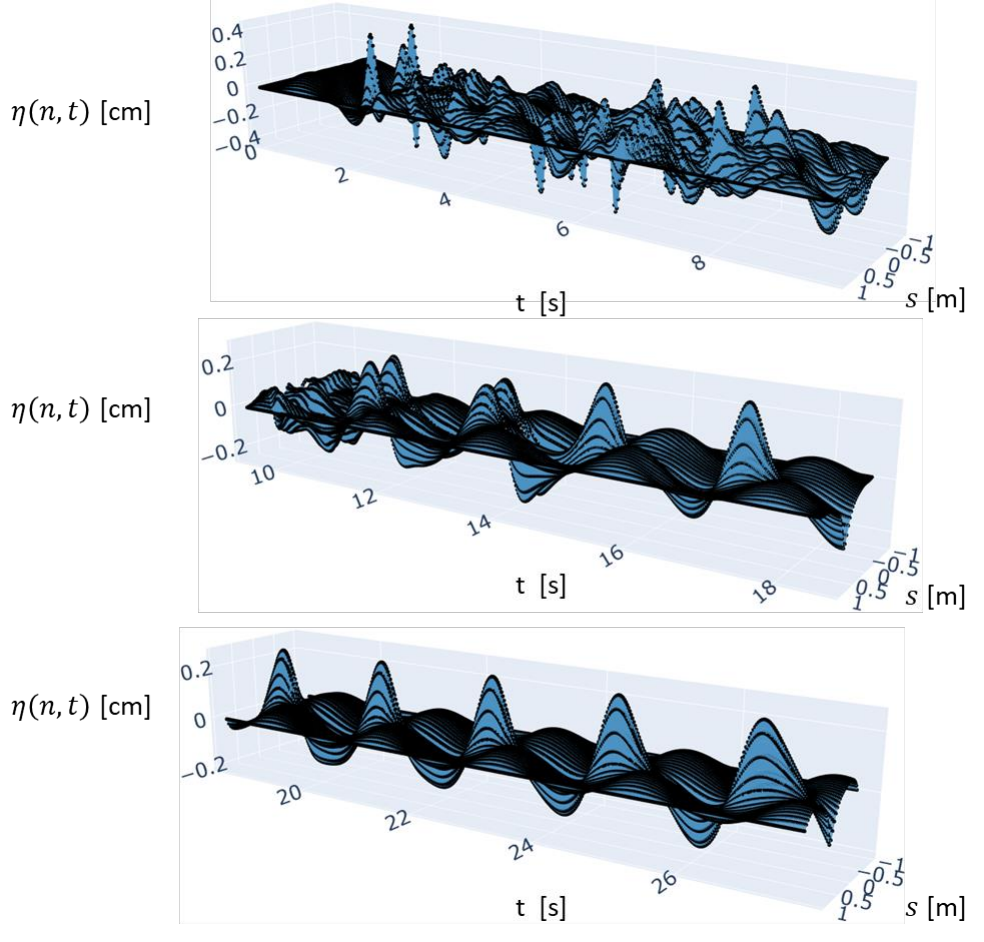}
\caption{Displacement $\eta$ of the oscillators as a function of time and of their position. Parameters used according to Tables \ref{tab2} and \ref{tab3}. Integration method and $\delta$ value, Radau and $0.1$ respectively. Black dots representing the positions of each oscillator}
\label{Fig10}       
\end{figure*}

Figure \ref{Fig10} illustrates the dynamic behavior of the oscillators after the stability condition is imposed. As observed, temporal stability, which characterizes soliton behavior, is replicated in the simulation. The boundary condition is respected throughout the integration ($\eta_0 = \eta_{N+1} = 0$), and, due to the choice of an odd number of oscillators, the largest displacements are found in the oscillator located at position $s = 0$.

\subsubsection{Allocation of kinetic energy}

The formation of solitons is being presented from the perspective of the oscillation amplitude along the chain. Although, intuitively, the allocation of kinetic energy occurs parallel to the location of highest amplitudes, a numerical analysis must be performed in order to validate this hypothesis.

The energy forms of the system transition between potential energy in the springs and kinetic energy due to the velocity of each mass. Simultaneously, part of the energy is dissipated by the damper while the system is fed back through the acceleration $\ddot{B}$. Such energies can be expressed by the following equations:

 \begin{equation}
        \begin{array}{c c c}

        \\ {E_k}_n(t)  = \frac{m}{2}\dot{\eta}_n^2(t), & {E_{pk_l}}_n(t) = \frac{k_l}{2}\eta^2(t), 
        \\
        \\ {E_{pk_{nll}}}_n(t) = \frac{k_{nl}}{4}\eta^4(t),  & {E_{pk_{c}}}_n(t) =  \frac{k_{c}}{2}\eta^2(t), 
        \\
        \\ {E_d}_n(t) = -c\int_{0}^{t} \dot{\eta}_n^2(t)\mathrm{d}t,

            \end{array} 
    \label{eq23}
\end{equation}where ${E_k}_n$, ${E_{pk_l}}n(t)$, ${E_{pk_{nl}}}n(t)$, ${E_{pk_{c}}}_n(t)$, and ${E_d}_n(t)$ represent the kinetic energy, potential energy of linear stiffness, potential energy of nonlinear stiffness, potential energy of coupling stiffness, and dissipated energy in each oscillator, respectively. The energy balance is given by the relation:

 \begin{equation}
    {E_B}_n = {E_c}_n + {E_{pk_l}}_n + {E_{pk_{nll}}}_n  + {E_{pk_{c}}}_n + {E_d}_n,
    \label{eq24}
\end{equation}where ${E_B}_n$ represents the energy injected at the base of the system.

\begin{figure*}[ht!]
  \centering
  \begin{tabular}{@{}c@{}}
    \includegraphics[scale=.7]{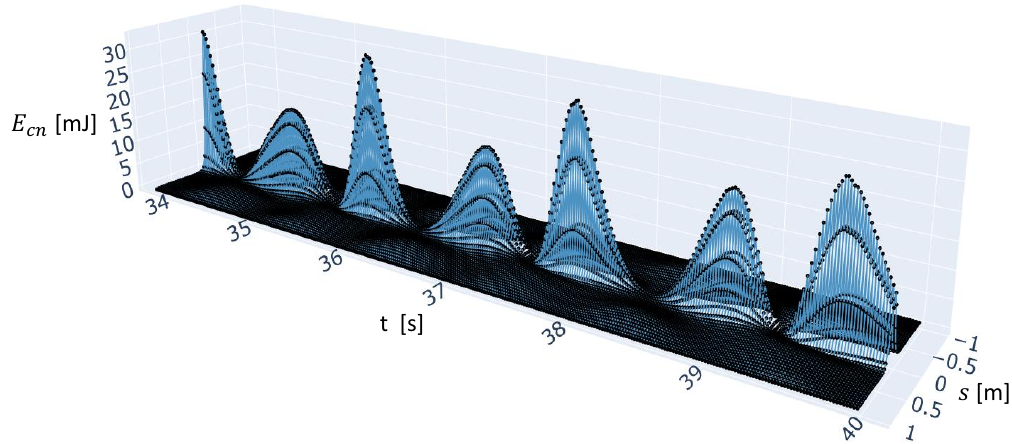} \\
    \small (a) 
  \end{tabular}
  
  \vspace{\floatsep}
  
  \begin{tabular}{@{}c@{}}
    \includegraphics[scale=.7]{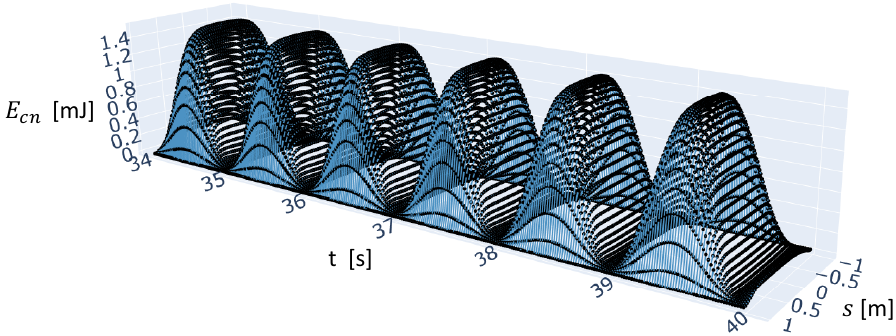} \\
    \small (b)
  \end{tabular}
  
  \caption{Allocation of kinetic energy of the oscillator chain depicted in Figure \ref{Fig10}. In (a), numerical analysis employing the design methodology. In (b), integration of Equation \ref{eq1} from a motionless configuration. Black dots representing the positions of each oscillator}
  \label{Fig11}
\end{figure*}

The description of the kinetic energy over time is shown in Figure \ref{Fig10} ($\delta = 0.1$). By comparing the solitonic dynamics in Figure \ref{Fig11}a with the non-solitonic dynamics in Figure \ref{Fig11}b, we can infer not only intuitively but also numerically that the oscillation energy migrates from the oscillators closer to the edges to the oscillators closer to the center of the structure, causing an energy localization.

\newpage

The oscillator positioned at $s = 0$, which exhibits the largest amplitude of oscillation, allocates the highest kinetic energy within the chain, approximately $33.03~mJ$ of maximum energy in steady-state under solitonic oscillation. In contrast, the same oscillator in the absence of localization, achieves a maximum kinetic energy in steady-state of approximately $1.53~mJ$. The difference between these two values is approximately $21.6$ times, confirming the property of solitons to concentrate their energy in localized regions of the space doamin.

\section{Concluding remarks}
\label{conclusion}
 
This paper presents a design methodology for reproducing and manipulating localized modes in chains of nonlinear damped oscillators. The present study explores the replication of stationary solitons as a strategy for obtain energy localization. The proposed methodology, by utilizing the NLSE as a design guide, holds potential for practical applications, once the replication of damped stationary NLSE remains challenging from a experimental standpoint when only external excitation is considered.

The strategy contributes to the field of energy localization in nonlinear lattices by presenting an analysis regarding the influence of the number of oscillators in the periodic structure; a numerical investigation of localized modes starting from an initial motionless configuration and a relationship between coupling stiffness and the allocation of oscillators in regions of interest in the spatial domain of NLSE.

Through the selection of the parameter $\delta$, the algorithm design is capable of allocating a desired number of oscillators in the region of maximum solitonic amplitude, enabling the designer to determine the intensity of the energy localization phenomenon. This procedure contributes to the understanding not only of achieving ILMs but also of innovative strategies to control them.

Overall, this research expands our understanding of mechanical solitons from a mechanical engineering perspective and lays the foundation for future studies in this area. Through the proposed design algorithm, experimental validations can be achieved, which opens up new possibilities for practical applications and advancements in innovative sensing technologies, dynamics analysis, and metamaterials.

\section*{Declaration of competing interest}

The authors declare that they have no known competing financial interests or personal relationships that could have appeared to influence the work reported in this paper. 

\section*{Acknowledgments}

We gratefully acknowledge support from EUR EIPHI program, Europe (Contract No. ANR 17-EURE-0002).

\end{document}